\documentclass{article}%
\usepackage{epsf} 
\usepackage{amsmath}
\usepackage{amssymb}
\usepackage{epsfig}
\usepackage{latexsym}
\usepackage{amsfonts}
\usepackage{graphicx}%
\usepackage{varioref}
\usepackage{ifthen}
\begin{document}
\parindent 0mm 
\setlength{\parskip}{\baselineskip} 
\thispagestyle{empty}
\pagenumbering{arabic} 
\setcounter{page}{0}
\mbox{ }
\rightline{UCT-TP-268/07}
\newline\rightline{July 2007}
\newline
\vspace{0.2cm}

\begin{center}
{\large {\bf Heavy-light quark pseudoscalar and vector mesons at finite temperature}}
{\LARGE \footnote{{\LARGE {\footnotesize Supported in part by FONDECYT 1051067, 7050125, and 1060653, by Centro de Estudios Subatomicos (Chile), and by NRF (South Africa).}}}}
\end{center}

\vspace{.1cm}
\begin{center}
{\bf Cesareo A. Dominguez}$^{(a)}$, {\bf Marcelo Loewe},$^{(b)}$, {\bf J. Cristobal Rojas},$^{(c)}$
\end{center}
\begin{center}
$^{(a)}$Centre for Theoretical Physics and Astrophysics\\[0pt]University of
Cape Town, Rondebosch 7700, South Africa

$^{(b)}$Facultad de F\'{i}sica, Pontificia Universidad Cat\'{o}lica de Chile, Casilla 306, Santiago 22, Chile

$^{(c)}$ Departamento de F\'{i}sica,  Universidad Cat\'{o}lica del Norte, Casilla 1280, Antofagasta, Chile
\end{center}
\vspace{0.3cm}
\begin{center}
\textbf{Abstract}
\end{center}
The temperature dependence of the mass, leptonic decay constant, and width of heavy-light quark peseudoscalar and vector mesons is obtained in the framework of thermal Hilbert moment QCD sum rules. The leptonic decay constant of both pseudoscalar and vector mesons decreases with increasing $T$, and vanishes at a critical temperature $T_c$, while the mesons develop a width which increases dramatically and diverges at $T_c$, where $T_c$ is the temperature for chiral-symmetry restoration/quark-gluon deconfinement. These results indicate the disappearance of hadrons from the spectral function,  which then becomes a smooth function of the energy.  This is interpreted as a signal for deconfinement at $T=T_c$. In contrast, the masses show little dependence on the temperature, except very close to $T_c$, where the pseudoscalar meson masses increase slightly by 10-20 \%, and the vector meson masses decrease by some 20-30 \%.


\newpage
\bigskip
\noindent
\section{Introduction}
\noindent
The thermal behaviour of hadronic Green's functions, obtainable in a variety of theoretical frameworks, plays a fundamental role in understanding the dynamics of the quark-gluon plasma. One such framework is that of  QCD sum rules \cite{QCDSR}, based on the Operator Product Expansion (OPE) of current correlators beyond perturbation theory, and on the notion of quark-hadron duality. The extension of this program to finite temperature was first discussed long ago in \cite{BOCH}. It is based on two basic assumptions, (a) that the OPE remains valid at $T \neq 0$, with  perturbative QCD (PQCD) and the vacuum condensates developing a temperature dependence, and (b) that the notion of quark-hadron duality also remains valid. Additional evidence, from solvable quantum field theory models, supporting these assumptions was provided later in \cite{CAD1}. Numerous applications of this technique have been made over the years \cite{VARIOUS}, leading to the following consistent picture. (i) With increasing temperature, particles that are stable at $T=0$  develop a non-zero  width, and resonances become broader, diverging at a critical temperature interpreted as the  deconfinement temperature ($T_c$). This width is a result of particle absorption in the thermal bath. (ii) The onset of the  continuum in hadronic spectral functions, traditionally accounted for by PQCD, decreases and approaches threshold near $T_c$. In other words, as $T \rightarrow T_c$ hadrons melt and disappear from the hadronic spectral functions, which become perfectly smooth. (iii) This scenario is further supported by results for hadronic and electromagnetic mean-squared radii, which also diverge at $T_c$ \cite{radii}. In addition, QCD sum rules in the axial-vector channel have provided (analytical) evidence for the near equality of the critical temperatures for deconfinement and chiral-symmetry restoration \cite{CAD2}. Regarding the temperature dependence of a hadronic mass, this parameter does not appear to be a relevant signal for deconfinement. Conceptually, given either the emergence or the broadening of an existing width, together with its divergence at $T_c$, the concept of mass looses its meaning. In practical applications,  in some cases the mass increases slightly with increasing $T$, and in others it decreases.\\
In this paper we use Hilbert moment QCD sum rules for heavy-light quark pseudoscalar and vector meson correlators to  determine the thermal behaviour of the hadronic masses, couplings, and widths. At $T=0$ this problem was discussed long ago in \cite{CAD_NP1}-\cite{BROAD}. While there are only four ground-state pseudoscalar heavy-light quark mesons in the spectrum ($D$, $D_s$, $B$, and $B_s$), and similarly for vector mesons, it is possible to determine the decay constants  for arbitrary meson masses in a self-consistent way \cite{CAD_NP2}. The result is that the leptonic decay constants obey a scaling law as a function of the meson mass. Since we are only interested  in the temperature behaviour of the hadronic parameters, we shall normalize all our results to those at $T=0$, thus obtaining a universal functional relation. We find that the meson masses are basically independent of $T$, except very close to $T_c$ where they increase slightly (pseudoscalars) by 10 - 20 \%, or decrease (vector mesons) by 20-30 \%. Here $T_c$ is the critical temperature for chiral-symmetry restoration/quark-gluon deconfinement. The leptonic decay constants decrease with increasing $T$, and vanish at the critical temperature. Pseudoscalar mesons  develop a non-zero hadronic width that increases with increasing $T$ and diverges at $T_c$, while vector meson widths exhibit a similar beahaviour. These results may be interpreted as providing (analytical) evidence for quark deconfinement at $T=T_c$.
\section{Pseudoscalar mesons}
We begin by defining the correlator of axial-vector divergences at finite temperature, i.e. the retarded Green's function

\begin{equation}
\psi_{5} (q^{2},T)   = i \, \int\; d^{4} \, x \; e^{i q x} \; \;\theta(x_0)\;
<<|[\partial^\mu A_{\mu}(x) \;, \; \partial^\nu A_{\nu}^{\dagger}(0)]|>> \;,
\end{equation}
where $\partial^\mu A_{\mu}(x) = m_Q :\bar{q}(x) \,i \, \gamma_{5}\, Q(x):\;$,  $q$ ($Q$) refers to the light (heavy) quark field, and $m_Q >> m_q$ is assumed.  The matrix element above is the Gibbs average


\begin{equation}
<< A \cdot B>> = \sum_n exp(-E_n/T) <n| A \cdot B|n> / Tr (exp(-H/T)) \;,
\end{equation}

where $|n>$ is any complete set of eigenstates of the (QCD) Hamiltonian. We use here the quark-gluon basis, which allows for a smooth extension of the QCD sum rule program to non-zero temperature \cite{CAD1}. At $T=0$ and to leading order in PQCD \cite{BROAD}

\begin{equation}
\frac{1}{\pi}\, Im \,\psi_5(x,0) = \; \frac{3}{8\,\pi^2}\; m_Q^4\;
\frac{(1 - x)^2}{x} \;,
\end{equation}

where $x \equiv m_Q^2/s$, with $s  \geq m_Q^2$, and $0 \leq x \leq 1$. At finite temperature there are two distinct contributions to the correlator, the so called  scattering term ($q^2$ space-like), and the annihilation term ($q^2$ time-like) \cite{BOCH}. After a straightforward calculation we find the former to be exponentially suppressed, so that it can be safely neglected, while the latter is given by

\begin{equation}
\frac{1}{\pi}\, Im \,\psi_5(x,T) = \frac{1}{\pi}\, Im \,\psi_5(x,0) \;  \; \Biggl\{1 - n_F\Bigl[\frac{\omega}{2\,T}\, (1 + x)\Bigr]
- n_F\Bigl[\frac{\omega}{2\,T}\, (1 - x)\Bigr]\Biggr\} \;,
\end{equation}

where $Im \,\psi_5(x,0)$ is given by Eq.(3), $n_F(z) = (1+e^z)^{-1}$ is the Fermi thermal function, and in the rest frame (${\bf q} =0$) of the thermal bath   $x= m_Q^2/\omega^2$. At temperatures below $T \simeq 200 \;\mbox{MeV}$ one can safely assume the heavy quark mass to be temperature independent \cite{mQT}. The first thermal function above is exponentially suppressed and can be safely neglected for temperatures of order $\cal{O}$$(100\,- 200\, \mbox{MeV})$ , but the second one does contribute near threshold.\\
Up to dimension $d=6$ the non-perturbative expansion of the correlator at $T=0$ is given by \cite{CAD_NP1},\cite{BROAD}

\begin{eqnarray}
\psi_5(q^2)|_{NP} &=& \frac{m_Q^2}{m_Q^2 - q^2}\,  C_4 <O_4>\,+
\frac{m_Q^3}{4}\, \frac{q^2}{(m_Q^2 - q^2)^3} \, C_5 <O_5>
+ \frac{m_Q^2}{6}  \nonumber \\ [.3cm]
&\times& \left[ \frac{2}{(m_Q^2 - q^2)^2} - \frac{m_Q^2}{(m_Q^2 - q^2)^3}-\frac{m_Q^4}{(m_Q^2 - q^2)^4}\right]
C_6<O_6>  ,
\end{eqnarray}

where 

\begin{equation}
C_4<O_4> = \frac{1}{12\, \pi} \, <\alpha_s \, G^2> \,- \,m_Q \,<\bar{q}\, q>,
\end{equation}

\begin{equation}
C_5<O_5> = < g_s \,\bar{q}\, i \,\sigma_{\mu \nu}\, G^a_{\mu \nu}\, \lambda^a \, q>
\equiv 2 \;M_0^2 \;<\bar{q}\, q> \;,
\end{equation}

\begin{equation}
C_6<O_6> = \pi \alpha_s <(\bar{q} \gamma_\mu \lambda^a\, q) \sum_{q} \bar{q} \gamma^\mu \lambda^a\, q>\phantom{\frac{1}{1}} \stackrel{VS}{\Longrightarrow} - \frac{16}{9} \,\pi \,\alpha_s\,\rho |<\bar{q} q>|^2 \;, 
\end{equation}

and $m_c(m_c) \simeq 1.3\, \mbox{GeV}$, $m_b(m_b) \simeq 4.2\, \mbox{GeV}$, $<\bar{q}q> \simeq (- 250 \, \mbox{MeV})^3$, $\phantom{\frac{1}{1}}$ $\phantom{\frac{1}{1}}$  $<\alpha_s G^2/12 \pi> \simeq \,0.003 - 0.006\, \mbox{GeV}^4$, $M_0^2 \simeq 0.4 - 0.6 \, \mbox {GeV}^2$, and $\rho \simeq 3 - 5$ accounts for deviations from vacuum saturation (VS). Use of these values in Hilbert moment sum rules reproduce the pseudoscalar meson masses at $T=0$. However, they will play no crucial role in our analysis as changes in these parameters would only affect the normalization at $T=0$.\\

For the light-quark condensate at finite  temperature we shall use the result of \cite{GATTO}, obtained in the framework of the composite operator formalism, valid for the whole range of temperatures:  $T = 0 - T_c$, where $T_c$ is the critical temperature for chiral symmetry restoration. There is lattice QCD evidence \cite{LATT} as well as analytical evidence \cite{CAD2} for this critical temperature to be (almost) the same as that for deconfinement. The ratio $R(T) =<<\bar{q}{q}>>/<\bar{q}{q}>$  from \cite{GATTO} as a function of $T/T_c$ is shown in Fig. 1 .\\

The low temperature expansion of the gluon condensate is proportional to the trace of the energy-momentum tensor, and it starts only at order $T^8$ \cite{BERN}, viz.
\begin{equation}
<<\frac{\alpha_s}{12\, \pi} G^2>>= <\frac{\alpha_s}{12\, \pi} G^2> -\, \frac{\alpha_s}{\pi} \,\frac{\pi^4}{405}\, \frac{N_F^2 (N_F^2-1)}{33- 2 N_F} (\ln \frac{\Lambda_p}{T} - 1) \frac{T^8}{f_\pi^4}\;,
\end{equation}
where $N_F$ is the number of quark flavours, and $\Lambda_p \approx 200 - 400 \;\mbox{MeV}$. To a good approximation this can be written as

\begin{equation}
<<\frac{\alpha_s}{12\, \pi} G^2>>= <\frac{\alpha_s}{12\, \pi} G^2> \Big[ 1 - \Big (\frac{T}{T_c}\Big)^8\Big] \;.
\end{equation}

\begin{figure}[ht]
\begin{center}
\includegraphics[width=\columnwidth]{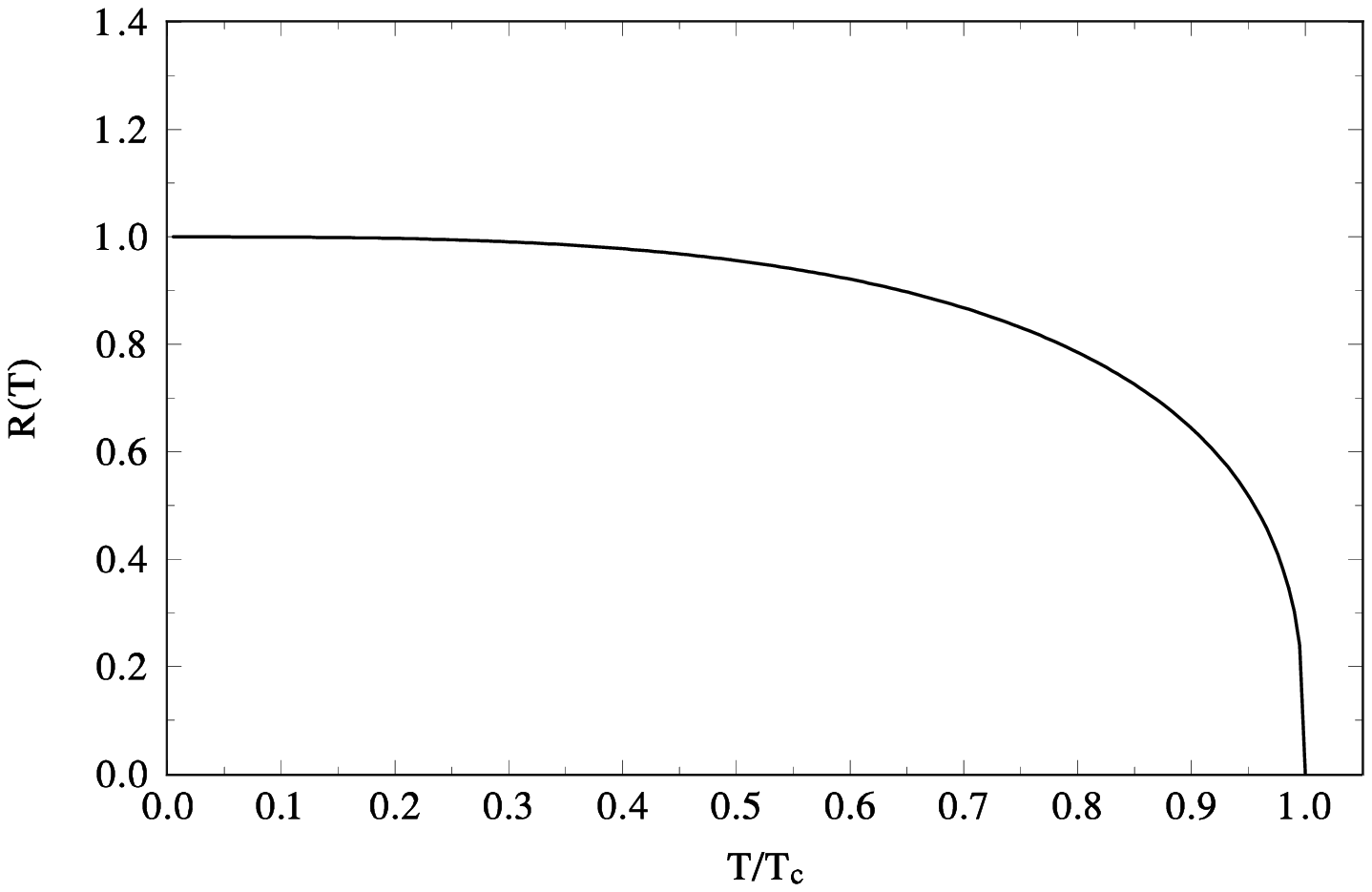}
\caption{The light-quark condensate ratio $R(T) = <<\bar{q}{q}>>/<\bar{q}{q}>$ as a function of $T/T_c$ from \cite{GATTO}.}
\end{center}
\end{figure}

Because of this $T$- dependence, the gluon condensate remains essentially constant up to temperatures very close to $T_c$. Hence, the thermal non-perturbative QCD correlator is basically driven by the quark condensate. Concerning the dimension $d=6$ condensate, it has been argued that the vacuum saturation approximation breaks down at finite temperature \cite{ELE}. This is based on the comparison between the slopes of the low temperature expansion (chiral perturbation theory) with and without assuming vacuum saturation. They are in fact numerically different. However, this result is only valid at very low temperatures ($T << f_\pi$); hence it cannot be extrapolated to $T \simeq T_c$. In fact, both the quark condensate and the four-quark condensate should vanish at the same temperature $T=T_c$. In any case, numerically, at temperatures of order $T \simeq 100 \, \mbox{MeV}$ the quark condensate dominates over the gluon condensate, the dimension $d=5$ condensate is comparable to $<<\bar{q}q>>$, and the dimension $d=6$ condensate is almost two orders of magnitude smaller. Hence, potential violations of vacuum saturation can be safely ignored. Finally, at finite temperature it is possible, in principle, to have non-zero  values of non-diagonal (Lorentz non-invariant) vacuum condensates. There is one example discussed in the literature  \cite{ELE2}  with enough detail to make a numerical estimate of their importance, and it refers to operators of spin-two (quark and gluon energy momentum tensors). The low temperature expansion of these terms starts at order $\cal{O}$$(T^4)$, in contrast to a $T^2$ dependence for the diagonal condensates. We find that at temperatures of order $T \simeq 100 \, \mbox{MeV}$ both non-diagonal condensates are three orders of magnitude smaller than the corresponding diagonal equivalents. We shall then ignore non-diagonal condensates in the sequel.\\
\begin{figure}[h]
\begin{center}
\includegraphics[width=\columnwidth]{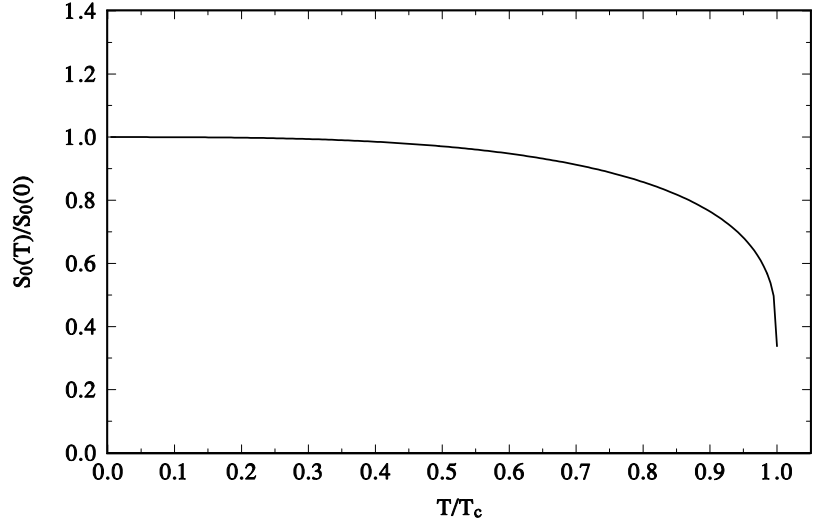}
\caption{The  ratio $s_0(T)/s_0(0)$, Eq.(14),  as a function of $T/T_c$ for $m_Q = m_c$.}
\end{center}
\end{figure}

Turning now to the hadronic sector, the spectral function at $T=0$ can be written as

\begin{equation}
\frac{1}{\pi}\, Im \,\psi_5(s)|_{HAD} = \; 2 \, f_P^2 \, M_P^4 \, \, \delta(s - M_P^2) \,+\, \theta(s - s_0) \, \frac{1}{\pi} Im \,\psi_5(s)|_{PQCD} \;,
\end{equation}

where $M_P$ and $f_P$  are the mass and leptonic decay constant of the pseudoscalar meson, and the continuum, starting at some threshold $s_0$,  is modeled by perturbative QCD. With this normalization, $f_\pi \simeq 93 \, \mbox{MeV}$.
Anticipating the pseudoscalar mesons  to develop a sizable width $\Gamma_P(T)$ at finite temperature (particle absorption in the thermal bath), and using a Breit-Wigner parametrization, the following replacement will be understood

\begin{equation}
\delta(s- M_P^2) \Longrightarrow const \; \frac{1}{(s-M_P^2)^2 + M_P^2 \Gamma_P^2}\; ,
\end{equation}

where the mass and width are $T-$dependent, and the constant is fixed by requiring equality of areas, e.g. if the integration is in the interval $(0 -\infty)$ then $ const = 2 M_P \Gamma_P/\pi$. 
\begin{figure}[h]
\begin{center}
\includegraphics[width=\columnwidth]{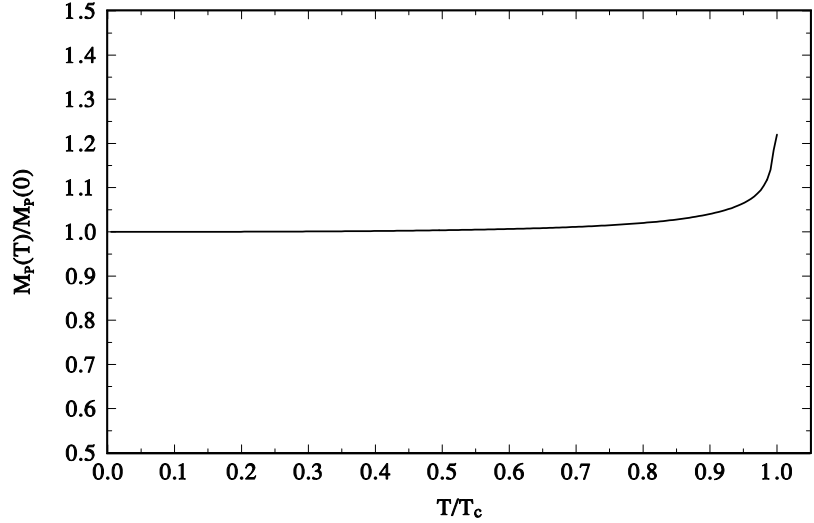}
\caption{The  ratio $M_P(T)/M_P(0)$ as a function of $T/T_c$.}
\end{center}
\end{figure}
The continuum threshold $s_0$ above also depends on temperature; to a very good approximation it scales  universally as the quark condensate \cite{VMDT}, i.e.

\begin{equation}
\frac{s_0(T)}{s_0(0)} \approx \frac{<<\bar{q}q>>}{<\bar{q}q>} \;,
\end{equation}

where $s_0(0)$ is clearly channel dependent. At the critical temperature we expect $s_0(T_c) = m_Q^2$, in which case Eq. (13) can be rewritten as

\begin{equation}
\frac{s_0(T)}{s_0(0)} = \frac{<<\bar{q}q>>}{<\bar{q}q>} \Biggl[ 1 - \frac{m_Q^2}{s_0(0)}\Biggr] + \frac{m_Q^2}{s_0(0)} \;,
\end{equation}
This is shown in Fig. 2 for the case $m_Q=m_c$ and $s_0(0) = 5\, \mbox{GeV}^2$; a qualitatively similar behaviour is obtained for $m_Q=m_b$ and $s_0(0) \simeq (1.1 - 1.3) M_B^2$.\\
\newpage
\begin{figure}[ht]
\begin{center}
\includegraphics[width=\columnwidth]{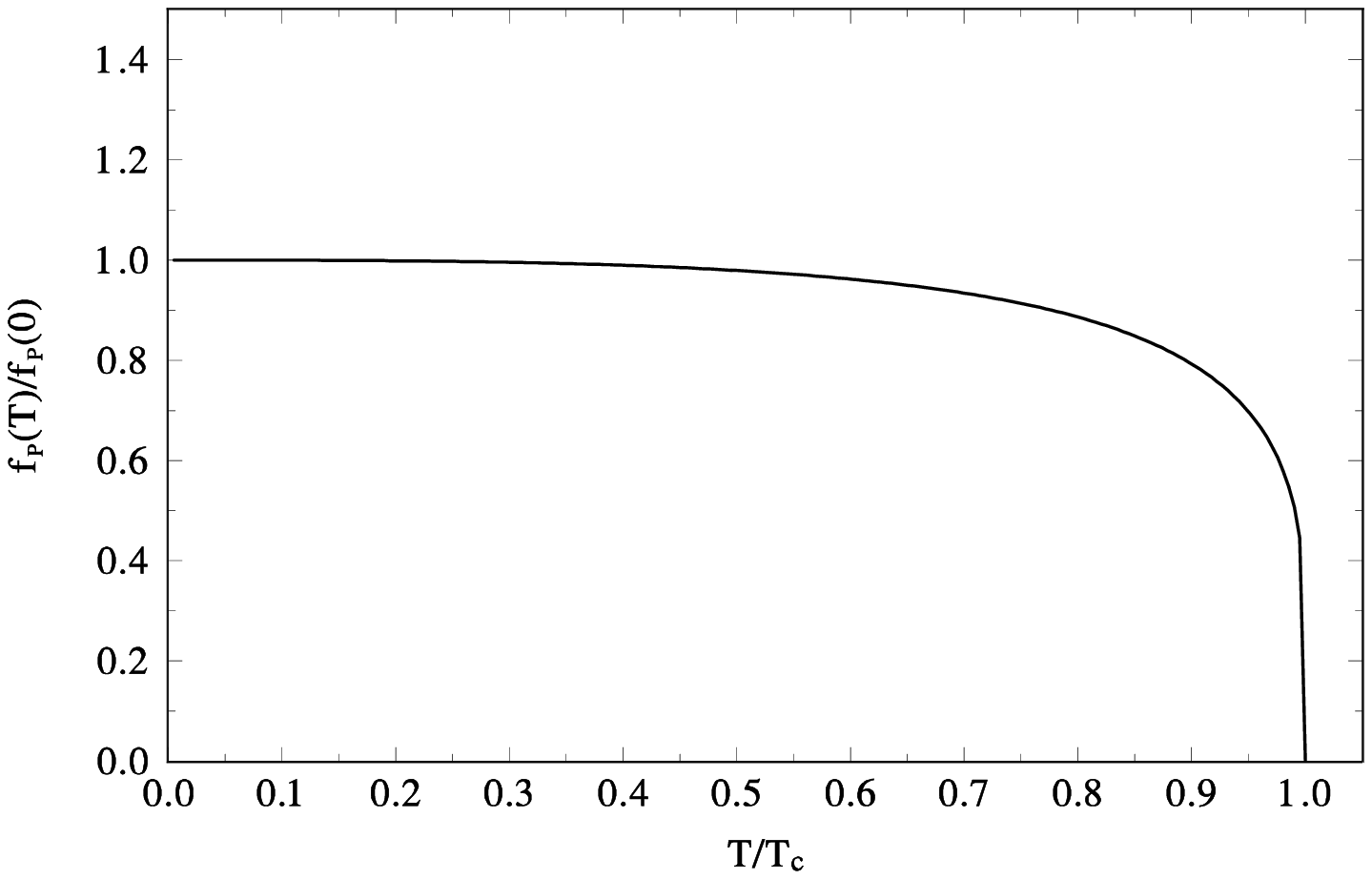}
\caption{The  ratio $f_P(T)/f_P(0)$ as a function of $T/T_c$.}
\end{center}
\end{figure}
The correlation function $\psi_5(q^2,T)$, Eq.(1), satisfies a twice subtracted dispersion relation. To eliminate the subtractions one can use Hilbert moments at $Q^2 \equiv - q^2 =0$, i.e.

\begin{equation}
\varphi^{(N)}(T) \equiv \frac{(-)^{N+1}}{(N+1)!}\, \Bigl(\frac{d}{dQ^2}\Bigr)^{N+1} \psi_5(Q^2,T)|_{Q^2=0}\, = \frac{1}{\pi}
\int_{m_Q^2}^{\infty} \; \frac{ds}{s^{N+2}}\, Im \,\psi_5(s,T)\; , 
\end{equation}

where $N = 1,2,...$. 
Invoking quark-hadron duality

\begin{equation}
\varphi^{(N)}(T)|_{HAD} =  \varphi^{(N)}(T)|_{QCD}
\;,
\end{equation}
\begin{figure}[ht]
\begin{center}
\includegraphics[width=\columnwidth]{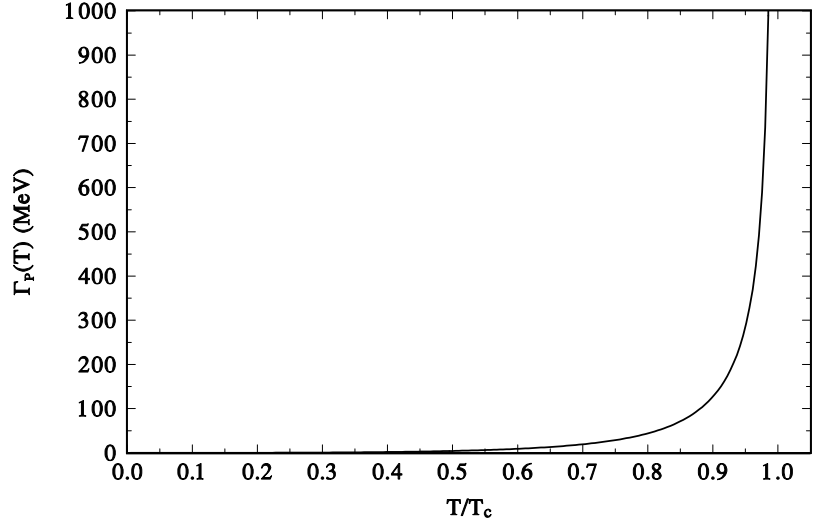}
\caption{The  width $\Gamma_P(T))$ as a function of $T/T_c$, with $\Gamma_P(0) = 0$.}
\end{center}
\end{figure}

and combining the continuum contribution in the hadronic spectral function with the PQCD piece of the QCD counterpart leads to the finite energy Hilbert moments

\begin{eqnarray}
\frac{1}{\pi}
\int_{0}^{s_0(T)} \frac{ds}{s^{N+2}}\, Im \,\psi_5(s,T)|_{POLE} &=& \frac{1}{\pi}
\int_{m_Q^2}^{s_0(T)} \frac{ds}{s^{N+2}}\, Im \,\psi_5(s,T)|_{PQCD} \nonumber \\ [.3cm]
&+& \varphi^{(N)}(T)|_{NP} \;,
\end{eqnarray}

where $Im \,\psi_5(s,T)|_{POLE}$ is given by the first term in Eq.(11) modified according to Eq.(12), the PQCD spectral function corresponds to Eq.(4), and 

\begin{eqnarray}
\varphi^{(N)}(T)|_{NP} &=& - \frac{m_Q <<\bar{q} q>>}{m_Q^{2N+2}} \left[ 1
- \frac{1}{12 \pi} \frac{<<\alpha_s G^2>>}{m_Q <<\bar{q}q>>} - \frac{1}{4}
(N+2)(N+1) \right. \nonumber \\ [.3cm]
&\times& \left.  \frac {M_0^2}{m_Q^2} - \frac{4}{81} (N+2) (N^2 + 10\, N + 9) \,\pi\, \alpha_s\, \rho\,
\frac{<<\bar{q}q>>}{m_Q^3} \right] \;.
\end{eqnarray}

Using the first three moments one obtains the temperature dependence of the mass, the leptonic decay constant, and the width. Results from this procedure are shown in Figs.3-5 for the charm case; in the case of beauty mesons, results are qualitatively similar.

\section{Vector mesons}
\noindent
We consider the correlator of the heavy-light quark vector current

\begin{eqnarray}
\Pi_{\mu\nu} (q^{2},T)   &=& i \, \int\; d^{4} \, x \; e^{i q x} \; \;\theta(x_0)\;
<<|[ V_{\mu}(x) \;, \; V_{\nu}^{\dagger}(0)]|>> \nonumber \\ [.3cm]
&=& -(g_{\mu\nu} q^2 - q_\mu q_\nu) \Pi^{(1)}(q^2,T) + q_\mu q_\nu \Pi^{(0)}(q^2,T) \; ,
\end{eqnarray}

where $V_\mu(x) = : \bar{q}(x) \gamma_\mu Q(x):$. In the sum rule analysis we shall use the function $- Q^2 \Pi^{(1)}(Q^2,T)$, which is free of kinematical singularities. A straightforward calculation gives

\begin{equation}
\frac{1}{\pi}\, Im \,\Pi^{(1)}(x,T) = \frac{1}{8 \pi^2}(1-x)^2 (2+x) \Bigl[1 - n_F\bigl(z_+\bigr)
- n_F\bigl(z_-\bigr)\Bigr] \;,
\end{equation}

where $z_{\pm} \equiv \frac{\omega}{2\,T}\, (1 \pm x)$. In the hadronic sector, we define the vector meson leptonic decay constant $f_V$ through

\begin{equation}
<0| V_\mu(0) | V(k)> = \sqrt{2}\; M_V \;f_V \;\epsilon_\mu \; ,
\end{equation}

so that the pole contribution to the hadronic spectral function is $2 f_V^2 \delta(s - M_V^2)$. At $T=0$ the vector meson $D^{*} (2010)$  has a very small width in the $keV$ range ($96 \pm 22 \,keV)$, which we expect to increase with increasing $T$, so that the replacement in Eq.(12) will be made.\\
\begin{figure}[h]
\begin{center}
\includegraphics[width=\columnwidth]{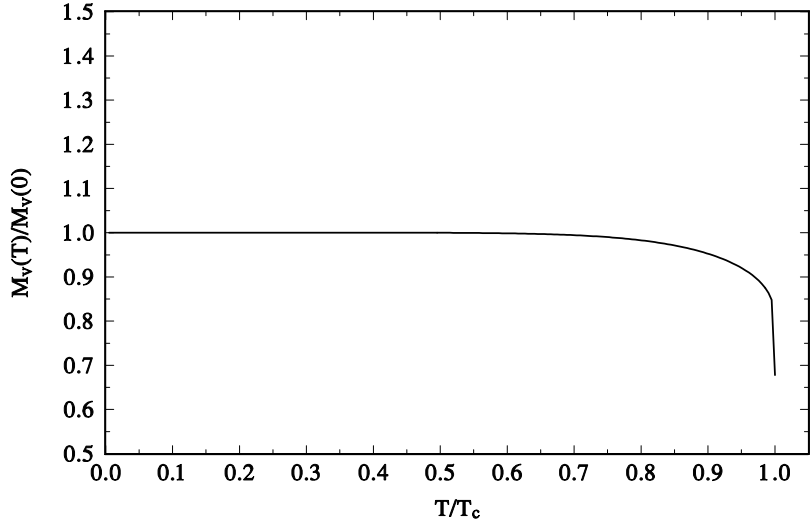}
\caption{The  ratio $M_V(T)/M_V(0)$ as a function of $T/T_c$.}
\end{center}
\end{figure}
\newpage
\begin{figure}[ht]
\begin{center}
\includegraphics[width=\columnwidth]{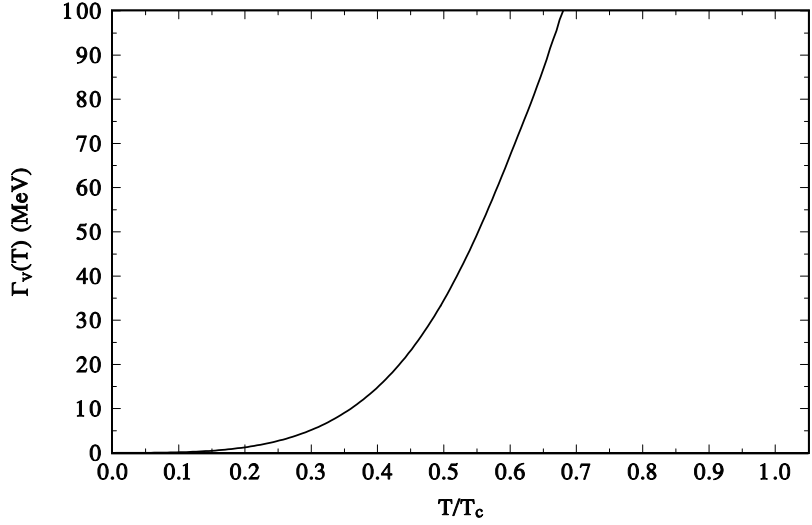}
\caption{The   width $\Gamma_V(T)$ as a function of $T/T_c$.}
\end{center}
\end{figure}
The Hilbert moments at $Q^2=0$ of the function $- Q^2 \Pi^{(1)}(Q^2,T)$ are given by

\begin{eqnarray}
\varphi^{(N)}(T) &\equiv& \frac{(-)^{N+1}}{(N+1)!}\, \Bigl(\frac{d}{dQ^2}\Bigr)^{N+1} \;[- Q^2 \Pi^{(1)}(Q^2,T)]|_{Q^2=0}\, \nonumber \\ [.3cm] &=& \frac{1}{\pi}
\int_{m_Q^2}^{\infty} \; \frac{ds}{s^{N+1}}\, Im \; \Pi^{(1)}(s,T)\; .
\end{eqnarray}

Following the same procedure as for the pseudoscalar mesons (see Eq. (17)), the sum rules become

\begin{eqnarray}
\frac{1}{\pi}
\int_{0}^{s_0(T)} \frac{ds}{s^{N+1}}\, Im \,\Pi^{(1)}(s,T)|_{POLE} &=& \frac{1}{\pi}
\int_{m_Q^2}^{s_0(T)} \frac{ds}{s^{N+1}}\, Im \,\Pi^{(1)}(s,T)|_{PQCD} \nonumber \\ [.3cm]
&+& \varphi^{(N)}(T)|_{NP} \;,
\end{eqnarray}

where $\varphi^{(N)}(T)|_{NP}$ is given by

\begin{eqnarray}
\varphi^{(N)}(T)|_{NP} &=& - \frac{m_Q <<\bar{q} q>>}{m_Q^{2N+4}} \Bigl[ 1
-  \frac{<<\alpha_s G^2>>}{12 \pi m_Q <<\bar{q}q>>} - \frac
{(N+2) (N+3)}{4}\Bigr. \nonumber \\ [.3cm]
&\times& \Bigl.  \frac {M_0^2}{m_Q^2} + \frac{4}{81} (N+2) (20\,+ N \,-N^2 ) \,\pi\, \alpha_s\, \rho\,
\frac{<<\bar{q}q>>}{m_Q^3} \Bigr] \;.
\end{eqnarray}

Using the first three Hilbert moments to find the temperature dependence of the hadronic parameters, we obtain for the mass and the width of $D^*(2010)$ the results shown in Figs.6-7. The behaviour of the vector-meson leptonic decay constant will not be shown, as it is essentially the same as that of the pseudoscalar-meson in Fig.4. Similar results are found for the case of the beauty vector meson $B^*$.

\section{Conclusions}
The thermal behaviour of pseudoscalar and vector meson decay constants, masses, and widths was obtained in the framework of Hilbert moment finite energy  QCD sum rules. This behaviour is basically determined by the thermal light quark condensate on the QCD sector, and by the T-dependent continuum threshold on the hadronic sector. Normalizing to values at $T=0$, and using the method of \cite{CAD_NP2} for arbitrary masses, there follows a universal  relation for the hadronic parameters as a function of $T/T_c$. Results show that the decay constants decrease with increasing temperature, vanishing at $T=T_c$, while the widths increase and diverge at the critical temperature. Such a behaviour provides (analytical) evidence for quark-gluon deconfinement, and is in qualitative agreement with corresponding results obtained in the light-quark sector. Finally, pseudoscalar meson masses increase slightly  with temperature by some 10 - 20 \%, while vector meson masses decrease by 20 - 30 \%. Given the dramatic emergence of monotonically increasing widths $\Gamma(T)$, there is little if any significance of this temperature behaviour of the masses, i.e. the relevant signals for deconfinement are the vanishing of the leptonic decay constants and the divergence of the widths at $T=T_c$.

\end{document}